\documentclass[aip,pop,reprint]{revtex4-1}

\usepackage{natbib}
\usepackage{amsmath,amssymb}  
\usepackage{bm} 
\usepackage[utf8]{inputenc} 
\usepackage{textcomp} 
\usepackage{tensor}
\usepackage{graphicx}
\usepackage{epstopdf}
\usepackage{pdfsync} 

\newcommand{\curl}{\nabla\times}

\newcommand{\diver}{\nabla\cdot}
\newcommand{\grad}{\nabla}

\newcommand{\EE}{\mathbf{E}}
\newcommand{\AAv}{\mathbf{A}}
\newcommand{\BB}{\mathbf{B}}

\newcommand{\QQ}{\mathbf{Q}}
\newcommand{\OO}{\mathbf{\Omega}}
\newcommand{\RR}{\mathbf{R}}
\newcommand{\PP}{\mathbf{P}}

\newcommand{\JJ}{\mathbf{J}}

\newcommand{\nt}{\tilde{n}}

\newcommand{\uu}{\mathbf{u}}

\newcommand{\dadb}[2]{\frac{\del #1}{\del #2}}

\newcommand{\Af}{Alfv\'{e}n\ }

\newcommand{\beqn}[1]{\begin{equation} \label{#1}}
\newcommand{\eeqn}{\end{equation}}
\newcommand{\beqna}[1]{\begin{eqnarray} \label{#1}}
\newcommand{\eeqna}{\end{eqnarray}}

\newcommand{\apj}{Astrophys. J.}

\newcommand{\del}{\partial}
\newcommand{\prl}{Phys. Rev. Lett.}

\begin{document}

\title{Relaxed States in Relativistic Multi-Fluid Plasmas}

\author{Jesse Pino}\email{pino@llnl.gov}
\author{Hui Li} \affiliation{Los Alamos National Laboratory, Los Alamos, NM 87545}

\author{Swadesh Mahajan}
\affiliation{Institute for Fusion Studies, University of Texas at Austin, Austin, TX 78712}


\date{\today}

\begin{abstract}
The evolution equations for a plasma comprising multiple species of charged fluids with relativistic bulk and thermal motion are derived. It is shown that a minimal fluid coupling model allows a natural casting of the evolution equations in terms of generalized vorticity which treats the fluid motion and electromagnetic fields equally. Equilibria can be found using a variational principle based on minimizing the total enstrophy subject to energy and helicity constraints.  A subset of these equilibria correspond to minimum energy. The equations for these states are presented with example solutions showing the structure of the relaxed states. 

\end{abstract}
\pacs{52.27.Ny,52.35.We,52.55.Dy}
\maketitle

\section{Introduction}
In a variety of astrophysical settings such as jets powered by Gamma Ray Bursts, supermassive black holes, or pulsars, energies can be such that the bulk motion of a fluid approaches the speed of light, and/or the thermal motion is comparable to the rest mass of the particles that make up that fluid. These cases necessitate a relativistic treatment of plasma motion. Previously,  Mahajan\cite{Mahajan:2003ud} showed that the dynamics of relativistic charged fluids could be treated using a minimal coupling formalism in direct analogy to the canonical momentum prescription of single particle dynamics. This formalism has been successfully used to describe both linear and nonlinear waves in relativistic pair plasmas of pulsar magnetospheres \cite{PhysRevE.81.026403,asenjo:122108}.  Presently, we demonstrate that there is a natural minimization principle associated with the minimal coupling formalism. When total energy and the helicity of multiple charged species are conserved, the application of this minimization principle to configurations leads to relaxed equilibrium states. We believe that these states give a better description of the physical system and can be used as more accurate starting points for subsequent dissipation and slow evolution. 
 
The outline of this paper is as follows. In Section 2, we present the formalism behind relativistic magnetofluids  and derive a minimization principle for finding equilibrium states. In Section 3, we present some examples of one dimensional solutions to the equilibrium equations derived from the minimization principle. Finally, we conclude by summarizing our results and discussing the applicability of these relaxed states to astrophysical phenomena, namely the `striped wind' of a pulsar nebula \cite{1990ApJ...349..538C}.

\section{Relativistic charged fluids}
\label{sec2}
We consider an isolated system of multiple fluids, and take the mass-energy to be small enough that we can neglect changes to the geometry of space-time. We thus take the Minkowski metric $g^{\mu\nu} = diag(-1,1,1,1)$.  The velocity 4-vector is $U^{\mu}= (\gamma c, \gamma \uu)$, where $\uu$ is the local 3-velocity of the each fluid species  and $\gamma=(1-u^2/c^2)^{-1/2}$.  Then $U^2\equiv U^{\mu}U_{\mu} = -c^2$.  We assume local Maxwellian closure, so all of the fluid quantities are implicitly functions of position (e.g. $n$, $T$, $p$), with $n$ the proper density of the fluid, and the pressure $p=n k T$. 

Each species (labeled $s$) satisfies mass conservation,
\beqn{mass}
\del_{\nu} \Gamma_s^{\nu} =\del_{\nu}(n_s U_s^{\nu})= 0,
\eeqn
and has the stress-energy tensor 
\beqn{stress}
T_s^{\mu\nu} = p_s g^{\mu\nu} + n_s m_s G(z_s) U^{\mu}U^{\nu},
\eeqn
where $n m c^2 G(z) = p+ne$ is the enthalpy (pressure plus internal energy) density of the fluid  \cite{deGroot1980},  and we assume some fluid closure that gives the enthalpy as a function temperature ($z = mc^2/kT$).  The results presented in this section are independent of the specific form of this enthalpy.  Each fluid has a charge $q_s$, coupling it to the electromagnetic field through the Lorentz force equation, 
\beqn{force1} 
\del_{\nu}T_s^{\mu\nu}  = (q\Gamma_{\nu})_s F^{\mu\nu} ,
\eeqn
where the E.M. field tensor $F^{\mu\nu}  = \del^{\mu}A^{\nu}-\del^{\nu}A^{\mu}$, with $A^{\mu} = (\phi, \bm{A})$ the 4-potential \cite{1998clel.book.....J}.  The E.M. field also obeys Maxwell's equations ($\mathcal{F}$ is the dual of $F$),
\[
\del_{\nu}F^{\mu\nu} = \frac{4\pi}{c}J^{\mu} \ , \ \ \del_{\nu}\mathcal{F}^{\mu\nu}=0,\]
 and has  stress-energy tensor
\beqn{EMstress}
T_{EM}^{\mu\nu} = \frac{1}{4\pi}\left(g^{\mu\alpha}F_{\alpha \beta}F^{\beta\nu} + \frac{1}{4}g^{\mu\nu}F^{\alpha\beta}F_{\alpha\beta}\right).
\eeqn
In order to close the system, we define the current $J^{\nu}=\sum_s q_s \Gamma_s^{\nu}  $ as the sum of the fluid currents, which is required so that the total stress-energy $T_{tot}^{\mu\nu}=T_{EM}^{\mu\nu}+\sum_s T_{s}^{\mu\nu}$ is divergence free ($\del_{\nu}T_{tot}^{\mu\nu}=0$).  

Expanding the force equation (\ref{force1}), we obtain (we suppress species labels except when needed for clarity): 
\beqn{forcefull}
 U^{\nu}\del_{\nu}(m G U^{\mu})+\frac{1}{n}\del^{\mu} p = q  F^{\mu\nu} U_{\nu}.
 \eeqn
 Using  $U^{\nu}\del_{\nu}= d/d \tau = \gamma d/dt = \gamma(\del/\del t+\uu\cdot\grad)$,  we can break this up into components:
 \beqn{timeforce}
 \frac{d}{dt} (m c^2 \gamma G) - \frac{1}{n \gamma}\frac{\del}{\del t} p = q  \uu\cdot \EE.
 \eeqn
 \beqn{spaceforce}
 \frac{d}{dt}  (m G \gamma \uu) +\frac{1}{n \gamma }\grad p = q (\EE + \uu\times\BB/c).
 \eeqn 
 This is the relativistic generalization of the Lorentz force equation, and we see that the effective mass is increased both by bulk motion $(\gamma)$ and thermal motion ($G$) \cite{1994PhRvL..73.1110B, PhysRevE.52.1968, 2002PhRvE..65d7402B}. 

If we define the entropy $\sigma$ through 
\beqn{entropydef}
T \del^{\nu} \sigma = mc^2 \del^{\nu} G - \frac{1}{n}\del^{\nu} p,
\eeqn
we can rewrite the evolution equation (\ref{forcefull}) for each species as
\beqn{UdotM}
 qU_{\mu}(F^{\mu\nu}+(c/q)S^{\mu\nu}) = q U_{\mu}M^{\mu\nu} = T \del^{\nu}\sigma  
\eeqn
where $S^{\mu\nu}=\del^{\mu}(m G U^{\nu})-\del^{\nu}(m G U^{\mu})$ is an antisymmetric tensor constructed using the temperature transformed momentum as a potential \cite{Lichnerowicz,1987ApJ...319..207B}.  This is the `Minimal Coupling Magnetofluid Unification' model described in Mahajan (2003) \cite{Mahajan:2003ud}.  The justification is that the combined potential $\Pi^{\nu}=A^{\nu}+(mc/q)GU^{\nu}$ can be seen as the canonical momentum of the fluid, just as one has $p\to p+qA$ for single particle motion. 
Since $M^{\mu\nu}$ is anti-symmetric, contracting equation (\ref{UdotM}) with $U_{\mu}$ gives $d\sigma/dt=0$,
the standard result of constant entropy along field lines. 

By using  fluid potential $mGU^{\nu}=(mcG\gamma,mG\gamma\uu)=(\chi,\PP)$ in place of $\phi$ and $\AAv$, we can define the  fields $\QQ=(-\del_t \PP/c -\grad \chi)$ and $\RR=\curl\PP$ as direct analogues to the electromagnetic fields $\EE$ and $\BB$. This allows us to expand (\ref{UdotM}) as
\beqn{entropygen}
q \uu\cdot\left(\EE + \frac{c}{q} \bm{Q}\right) =\frac{T}{\gamma} \frac{\del \sigma}{\del t}
 \eeqn
\beqn{evolution}
 \frac{\del}{\del t} \left( \AAv+\frac{c}{q}\PP \right) =
 \uu \times \OO - \grad \Phi +\frac{T}{q \gamma}\grad \sigma,
 \eeqn
 where we have defined the total species vorticity $\OO=\curl[\AAv+(c/q)\PP]=\BB+(c/q)\RR$, and potential $\Phi=\phi+(mc^2/q)G\gamma$. Equation (\ref{entropygen}) states that entropy is generated locally only if the fluid flow is not perpendicular to the sum of the electric field and the inertial analogue of the electric field. Equation  (\ref{evolution}) describes the evolution of the magnetic field and fluid momentum, including both gradient and inductive forces. Taking its curl, we obtain the generalized vorticity evolution equation for each species:
 \beqn{vorticity}
 \frac{\del \OO_s}{\del t} = \curl (\uu_s \times \OO_s)+\grad\left(\frac{T}{q\gamma}\right)\times\grad \sigma.
 \eeqn
 The first two terms in equation (\ref{vorticity}) state that the generalized vorticity is `frozen-in' to the flow. The last term describes vorticity generation due to the relativistic extension of the baroclinic term.  To make contact with the familiar MHD equations, we take a massless barotropic electron fluid with $\uu$ as the bulk plasma flow. We then recover the ideal MHD induction equation $\del_t \BB = \curl(\uu\times\BB)$, and the entropy generation rate $T \del_t \sigma = \JJ\cdot \EE$  . 

\section{Minimization principle}
We now seek equilibrium configurations of systems with multiple species. The presence of dissipation will allow ideally conserved quantities to change over time. In general, the conserved with the highest order spatial derivatives will change the fastest, and those with lower order derivatives can then be treated as constraints in a variational principle with the most susceptible quantity as the target of minimization  \cite{OS93}.  A standard example in MHD is the minimization of magnetic energy $(B^2)$, subject to the constraint of constant magnetic helicity $(\bm{A}\cdot\BB \sim B^2/k)$: \beqn{varprincE}
\delta(E-\alpha h)=0 .\eeqn
This results in the well-known Taylor state $\curl\BB=\alpha \BB$  \cite{Taylor74}. 
Within the current framework, species helicity and total energy are both ideally conserved variables under suitable assumptions (see Appendix). However, the kinetic terms cause the total helicity to include higher-order spatial derivatives than the total energy, making it more fragile to dissipation. Thus the variational principle (\ref{varprincE}) is mathematically ill-posed \cite{2002PhRvL..88i5001Y}. Simply minimizing helicity subject to constant energy is not possible either, since in general the helicity is not positive-definite and often not bounded from below.  Instead, we must find a coercive functional $N$ which is fragile to dissipation but has a quadratic form. We can then minimize $N$ while keeping both species helicity and total energy as constraints:
\beqn{fullminimize}
\delta\left(N-\mu E - \sum_{s}\alpha_{s} h_{s}\right) = 0.
\eeqn 
 The resulting states can then be compared to the equilibrium conditions to make sure they are physical, and we can further seek a subset of these relaxed states which correspond to minimum energy.  

We begin by constructing a canonical stress tensor for each species by replacing $F$ in equation (\ref{EMstress}) with the species canonical tensor $M^{\mu\nu}=F^{\mu\nu}+ (c/q)S^{\mu\nu}$:\beqn{tcanon}
\mathcal{T}_s^{\mu\nu} =
\frac{1}{4\pi} \left(g^{\mu\alpha}M_{\alpha \beta}M^{\beta\nu} + \frac{1}{4}g^{\mu\nu}M^{\alpha\beta}M_{\alpha\beta}\right)_s.
\eeqn
 Assuming that the system is isolated and that the fields vanish at the boundaries, the integral of the time-time component of this tensor will be an effective Hamiltonian for this system \cite{Holm:1989xr}. Summing over species, we see that this is equivalent to the sum of the species total (magnetic plus fluid) enstrophies 
 \beqna{enstrophy1}
 N&=&\sum_{s}  \frac{1}{8 \pi}\int\left[ \left|\grad\Phi\right|^2 + |\OO_s|^2\right] d^3 x .
 \eeqna 
The enstrophy is the most fragile to dissipation, since it contains the square of the curl of the fluid momentum ($\bm{R}$ terms in $\bm{\Omega}$). This makes it a proper target functional for minimization. We will minimize this with constraints of the total energy and species helicities, by varying the potentials $\phi$ and $\AAv$ and the momentum $\PP_{s}$.  Note that we do not independently vary $\chi$, the time component of $GU^{\mu}$, since at fixed temperature, 
\[
\delta\chi =   \delta (c G \gamma) =  \delta (\sqrt{c^2 G^2+P^2}) = \frac{\PP}{\chi}\cdot \delta\PP = \frac{\uu}{c}\cdot\delta\PP.
\]
The variation of the enstrophy is:
\beqna{deltaN}
\delta N = \sum_s\int \frac{1}{4\pi}\left\{ -\left[\ \nabla^2\Phi_s \right] \delta\phi +\left[ \curl \OO_s\right]\cdot \delta \AAv  \right. \nonumber \\
 \left.+ \frac{m_sc}{q_s}\left[\curl \OO_s - \frac{\uu_s}{c}\nabla^2\Phi_s \right]\cdot\delta\PP_s  \right\} d^3 x.
\eeqna

The total energy is the sum of electromagnetic and kinetic energies
\beqna{totalE}
E &=& \int T^{00} d^3 x= E_{EM}+E_s \\
&=& \int \frac{\EE^2+\BB^2}{8 \pi} d^3 x + \sum_{s} \int \left( m n c^2  \gamma^2  - p \right)_s d^3 x ,\nonumber
\eeqna
which leads to the variations:
\beqn{deltaEEM}
\delta E_{EM}= \int \frac{1}{4\pi}\left\{(-\nabla^2\phi)\delta\phi + (\curl\BB)\cdot\delta\AAv\right\}d^3 x,
\eeqn
\beqna{deltaEs}
\delta E_s &=& \delta\int n_s m_s c^2  G_s \gamma^2_s\ d^3 x \nonumber \\
&=& \int  n_s m_s \delta\left( \frac{c^2 G_s^2+\PP_s^2}{G_s}\right)\ d^3 x \nonumber \\
&=& \int 2 n_s m_s \gamma_s\uu_s \cdot \delta\PP_s\ d^3 x  , 
\eeqna

We define the generalized helicity in the Appendix, as 
\beqn{heldef1} h = \int [\BB+(c/q)\bm{R}]\cdot[\bm{A}+(c/q)\PP] d^3 x, \eeqn
which contains the fluid helicity, magnetic helicity and cross helicity terms. 
The variation of this helicity gives
\beqn{deltaH}
\delta h_s = \int\left\{ \OO_s \cdot \delta\AAv + \frac{c}{q_s} \OO_s \cdot \delta\PP_s \right\} d^3 x .\eeqn 

Plugging these variations into the minimization equation (\ref{fullminimize}) and considering each variation independently results in the set of equations:
\beqna{vareqns}
\sum_s \nabla^2\Phi_s -\mu \nabla^2 \phi =0  ,
\label{phieqn1} \\
\sum_s\left[\frac{1}{4\pi}\curl\OO_s -\alpha_s \OO_s\right] - \frac{\mu}{4\pi} \curl \BB  =0,
\label{Aeqn1}\\
\frac{1}{4\pi}\left[\curl \OO_s -\frac{\uu_s}{c} \nabla^2\Phi_s \right]-2 \mu n_s q_s \gamma_s \frac{\uu_s}{c}
\label{Pseqn1}\\
 -\alpha_s\OO_s =0.\nonumber
\eeqna
To ensure that these solutions are physical, (\ref{phieqn1}) and (\ref{Aeqn1}) must be equivalent to Poisson's and Ampere's equations, which requires
\beqn{phieqn2}
\sum_s \nabla^2\Phi_s= - 4 \pi \mu \sum_s n_s q_s \gamma_s .
\eeqn  
Substituting (\ref{Pseqn1}) into (\ref{Aeqn1}), 
\beqn{Aeqn2}
\frac{\mu}{4\pi} \curl \BB= \sum_s\left[\frac{1}{4\pi} \nabla^2\Phi_s+ 2 \mu n_s q_s \right] \frac{\uu_s}{c} .
\eeqn 
Thus each species component of (\ref{phieqn2}) must hold independently, viz. 
\beqn{phieqn3}
\frac{1}{\mu}\nabla^2\left(\phi+ \frac{m_s}{q_s}\chi_s\right) = - 4 \pi  n_s q_s \gamma_s .
\eeqn
The states with minimum enstrophy satisfy the relation 
 \beqn{Pseqn2}
 \frac{1}{4\pi} \curl \OO_s  -\alpha_s\OO_s = \mu \frac{n_s q_s \gamma_s }{c}\uu_s.
 \eeqn
 
 The subset of minimum energy states can be found by taking the limit $\mu,\alpha_s\to \infty$, keeping $\alpha_s/\mu = - 1/ \lambda_s$ constant.  This formally reproduces the energy minimization 
 $ \delta \left(E - \sum_s h_s/ \lambda_s\right) =0$, 
 however, by using the states constructed from (\ref{fullminimize}), we arrive at the relaxed states via a well-posed minimization problem. In order for both Ampere's law and (\ref{phieqn3}) to still be satisfied in the limit of large $\mu$, we must have 
\beqn{phieqn4}
\nabla^2\left(\phi+ \frac{m_s}{q_s}\chi_s\right) \to 0. 
\eeqn
For a system with $\grad(\phi+ \frac{m_s}{q_s}\chi_s)=0$ on the boundaries, this implies that 
\beqn{relaxedstate1}
\gamma_s G_s + \frac{q_s}{m_s c^2}\phi = \gamma_{max,s} = const, 
\eeqn
which means that the gradient forces in the evolution equation (\ref{evolution}) vanish. 
Since $G,\gamma\ge1$, $ mc^2\gamma_{max}$ can be interpreted as the total energy available to the fluid species, to be distributed among thermal, bulk motion, and electrostatic energies. 

The relaxed fluid states further satisfy the relation
\beqn{relaxedstate2}
  \curl (G \gamma \uu)_s=  \lambda_s \frac{n_s q_s^2 \gamma_s }{m_s c^2}\uu_s-\frac{q_s}{m_s c}\BB .
 \eeqn
This condition was given previously by Els\"{a}sser and Popel \cite{1996PhPl....3..482E} in the case of non-relativistic temperatures, without justification for it being a minimum energy state. This equilibrium condition introduces a natural length scale of the collisionless skin depth. To show this more concretely, we take the case of an electron-positron pair plasma. Normalizing length to some system size $L$, velocity to $c$, density to $n_0$, and magnetic field to $B_0= \sqrt{2 \pi m_e n_0 c^2}$, 
Equations (\ref{Aeqn2}) and (\ref{relaxedstate2}) become
 \begin{eqnarray}
 \epsilon \curl \PP_+&=&  \epsilon^{-1} \hat{\lambda}_+ \frac{n_+}{G_+} \PP_+  -\frac{ \BB}{2} ,\label{eeplus} \\
\epsilon\curl \PP_-&=& \epsilon^{-1} \hat{\lambda}_- \frac{n_-}{G_-} \PP_-  +\frac{\BB}{2}, \label{eeminus} \\
\epsilon \curl \BB &=& \frac{n_+}{G_+} \PP_+ - \frac{n_-}{G_-} \PP_-,\label{eeAmpere}
\eeqna
where $\hat{\lambda}=\lambda /8\pi L $, and $\epsilon= \lambda_{pe}/L = \sqrt{mc^2/8 \pi n_0 q^2}/L$ is the ratio of the pair skin depth to the system size.  $\PP_+=G_+\gamma_+\uu_+$ is the momentum of the positrons, and minus sign subscripts denote electron quantities. Previous work by Iqbal \emph{et al.} \cite{iqbal:032905} examined the limit of isotropic equal temperatures and densities, with quasineutrality and nonrelativistic bulk flow ($\gamma\sim 1$).  They showed that $\BB$ could be expressed as the sum of three Beltrami flows ($\curl \mathbf{F}_j = \mu_j \mathbf{F}_j)$, with different scale lengths $\mu_j^{-1}$. In appropriate limits, one of the eigenvalues $\mu$ can become very small, leading to structures on a scale much larger than the skin depth, making the solutions relevant to modelling of astrophysical jets. Indeed, if we assume that the spatial variation of the fields is only on the large scale, to first order we recover the MHD relaxed ``Taylor state'' $\curl \BB = \alpha \BB$, with $\alpha=(\hat{\lambda}_+^{-1}+\hat{\lambda}_-^{-1})/2$. However, the presence of the small parameter $\epsilon$ multiplying the curl operator means that the multi-fluid equations represent a singular perturbation on MHD dynamics, and that the small scale structure cannot be fully ignored.  In general, the relations between $G$, $\gamma$ and $n$ make the equilibrium equations (\ref{eeplus}-\ref{eeAmpere}) highly nonlinear, and solutions must be found numerically. In the following section, we examine the class of one dimensional solutions in various simplifying cases.  

\section{One Dimensional Solutions}
\label{examples}
In order to explore some configurations of the relaxed states, we will now restrict to the class of solutions in one dimension. We take a Cartesian coordinate system with all quantities varying in $x$ only.   We further consider those simple cases where the enthalpy, temperature and density of the two species are equal $(G_+=G_-=G,\ T_+=T_-=T,\ n_+=n_-=n)$. This in turn implies that $\lambda_+^2=\lambda_-^2$ (We drop the hat notation of the previous section). For definiteness, we take the relativistic Maxwellian distribution \cite{Synge57,deGroot1980,1989cup..book.....A}. The enthalpy then takes the form  $G(z)=K_3(z)/K_2(z)$, where $K_i$ is the MacDonald function of order $i$ and $z=mc^2/k_B T$. We also take the entropy to be constant within the region of interest so the density takes the form $n = n_0 (K_2(z)/z)$. This assumption can later be relaxed (see the following discussion). 

\subsection{Perpendicular current}
\label{reversal}
We first look for an analogue of the 1-D Harris-Hoh sheet \cite{Harris1962,1966PhFl....9..277H}, with  $\JJ \perp \BB$. This implies that the direction of B does not change, and we are free to choose $\BB=B(x) \hat{z}$, and $\JJ=J(x)\hat{y}=-B' \hat{y}$. We could take in addition a constant $J_{x0}$, but then the current would not vanish at the boundaries. Also note that the 1-D assumption precludes the possibility of a constant $B_{x0}$, since equations (\ref{eeplus}) and (\ref{eeminus}) would demand nonzero curl of $\PP$ in the x-direction.  The current equation (\ref{eeAmpere}) then becomes 
  \beqn{Beqn1}
\epsilon \dadb{B}{x} \hat{y} = - \frac{n}{G}  (\PP_+ - \PP_-) , 
  \eeqn
 implying that $P_{x-} = P_{x+}$, and $P_{z-} = P_{z+}$. 
 
 Crossing (\ref{eeplus}) and (\ref {eeminus}) with $\BB$, we obtain 
\beqna{perpplus} \epsilon\grad(\PP_+\cdot\BB) = \epsilon(\BB\cdot\grad)\PP_+ + \epsilon(\PP_+\cdot \grad) \BB \nonumber \\ 
+ \epsilon^{-1}\lambda_+ \frac{n}{G}( \BB\times \PP_+) 
+ \epsilon\PP_+  \times (\curl\BB) ,\eeqna
\beqna{perpminus} \epsilon\grad(\PP_-\cdot\BB) = \epsilon (\BB\cdot\grad)\PP_- + \epsilon(\PP_-\cdot \grad) \BB \nonumber \\ -\epsilon^{-1}\lambda_- \frac{n}{G} (\BB\times \PP_-) + \epsilon\PP_- \times (\curl\BB).\eeqna
The first term on the right implies that $P_x=0$ for both species. Then by inspection, the above equations are equivalent only if we take $\lambda_+=- \lambda_-=\lambda$, and  $\PP_+=  \PP_{\|}+\PP_{\perp} =P_z  \hat{z}+P_y \hat{y}, $ and $\PP_-= \PP_{\|}-\PP_{\perp}= P_z  \hat{z}-P_y \hat{y}. $ Thus there is a net current perpendicular to the magnetic field as well as net momentum parallel to the field. The current equation is then written
 \beqn{Beqn2}
\epsilon \dadb{B}{x} = - 2\frac{n}{G} P_y.
  \eeqn

Equations (\ref{perpplus}) and (\ref{perpminus}) become 
\beqn{ppareqn1}
  \epsilon\dadb{}{x}(P_z B) = - \frac{n}{G}\left(\epsilon^{-1}\lambda  B + 2 P_z  \right)P_y\eeqn
 Using (\ref{Beqn2})  allows us to write:
\[ \epsilon B\dadb{P_z}{x}  -2 \frac{n}{G}  P_z  P_y = -\frac{n}{G}\left(\epsilon^{-1}\lambda  B + 2 P_z  \right)P_y, \] or,
  \beqn{Pzeqn1}
  \epsilon \dadb{P_z}{x} = -\epsilon^{-1} \frac{n}{G}\lambda P_y. \eeqn
 
  Dotting  (\ref{eeplus}) and (\ref {eeminus}) with $\BB$ gives 
  \[ \epsilon \diver (\PP_{\perp}\times \BB )  = \epsilon^{-1} \lambda \frac{n}{G}\BB\cdot \PP -\frac{B^2}{2} - 2 \frac{n}{G} P_{\perp}^2 ,\]
  which in 1-D reduces to 
 \[
   \epsilon \dadb{}{x} (P_y B)  = \epsilon^{-1} \lambda \frac{n}{G}P_z B  -\frac{B^2}{2} -  2 \frac{n}{G} P_y^2  \]
   Using (\ref{Beqn2}), we get
 \beqn{Pyeqn1}
  \epsilon \dadb{P_y}{x}   = \epsilon^{-1} \lambda \frac{n}{G}P_z   -\frac{B}{2}    \eeqn 
 
 We have a set of three equations (\ref{Beqn2}), (\ref{Pzeqn1}), and (\ref{Pyeqn1}). In order to close these equations, we take the adiabatic distribution where $\gamma G=\gamma_{max}$. Then we are able to write $G^2 = \gamma_{max}^2-P^2$. The temperature and density can be solved for using the relativistic Maxwellian distribution described above. The nonlinearity inherent in this prescription necessitates numerical integration to find solutions. As an example, in Figure \ref{fig1}, we plot a 1-D equilibrium solution for a Maxwellian pair plasma containing a magnetic field reversal similar to a Harris sheet.
 \begin{figure}[ht]
   \centering
   \includegraphics[width=.45 \textwidth]{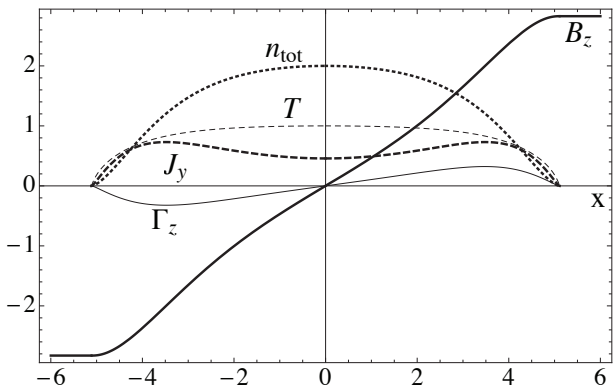} 
   \caption{Example of a pair-plasma equilibrium state in a 1-D $B_z$ field-reversal configuration, with  $\lambda_+=-\lambda_-=1$. Perpendicular to the field, the electrons and positrons stream in opposite directions, generating a current $J_y$. Along the field, both species flow together, creating a parallel flux $\Gamma_z=n \gamma u_z$. The adiabatic condition leads to peaked density ($n_{tot}=n_++n_-$) and temperature profiles, with points of vanishing density, beyond which the magnetic field is constant. Distance is normalized to the skin depth, and $n=n_0, k_B T=mc^2$ at $x=0$. }
   \label{fig1}
\end{figure}
 The electrons and positrons have equal temperatures and densities, but opposite eigenvalues $\lambda_+=-\lambda_-=1$. The adiabatic condition (\ref{relaxedstate1}) leads to a compact area of nonzero density and current, outside of which the magnetic field is constant.  The finite helicity of the system allows for bulk flow along the magnetic field. The two-fluid nature of the equations leads to field reversal on the scale of the skin depth, suggesting application to magnetic reconnection studies and shock structure.  The accessibility and stability of such nonlinear multi-fluid equilibria will be explored in forthcoming work. 

 \subsubsection{Osciliatory and Constant Solutions}
 \label{constant}
 Differentiating eq (\ref{Pyeqn1}), we have 
 \[ \epsilon^2 P_y''   =  \lambda \left(\frac{n}{G}\right)' P_z+ \epsilon^{-1} \lambda \frac{n}{G}\epsilon P_z'    -\epsilon \frac{B'}{2}\] 
 \beqn{pypp}
 \epsilon^2 P_y''   =  \lambda \left(\frac{n}{G}\right)' P_z- \left[ \epsilon^{-2} \lambda^2 \left(\frac{n}{G}\right)   - 1\right] \left(\frac{n}{G}\right)P_y \eeqn
 If the variation in $n/G$ is small, then the solution is oscillatory when 
 \[ \lambda   >  \epsilon \sqrt{G/n}= \frac{1}{L}\frac{\sqrt{m G}c}{\sqrt{8 \pi m n_0 \nt q^2}} =\frac{c/\omega_{p,eff}}{L },\]
 where $\omega_{p,eff}$ is the effective plasma frequency including thermal effects.   

 If $ \lambda   =  \epsilon \sqrt{G/n}$, and $P_y=0,$ there is a solution with constant $B, P_z, n,$ and $T$. From equation (\ref{Pyeqn1}), we see this requires $P_z=\sqrt{G/n}B/2$, or the total flow velocity (electrons plus positrons) is $u_z = B/\sqrt{nG\gamma^2}$. That is, the magnetic field is constant with flow along field lines at a relativistically modified \Af speed. 
   \subsubsection{Pressure Balance}
 \label{pressurebalance}
Examining Equations (\ref{Beqn2}), (\ref{Pzeqn1}) and (\ref{Pyeqn1}), we see that we have
 \beqn{b2eqp2}
\dadb{B^2}{x} =4\frac{n}{G} \dadb{P^2}{x}.
\eeqn
If we take the adiabatic distribution $\gamma G=\gamma_{max}$,  we are able to write $G^2 = \gamma_{max}^2-P^2$. At $x=0$, we take $B=B_0$, $T=T_0$, and $n=n_0$. The boundary $x_1$ is then defined as the point where the density vanishes and $P=P_{max}=\sqrt{ \gamma_{max}^2-1}$.  For $x>x_1$, the magnetic field is constant, $B=B(x_1)=B_1$. Then equation  (\ref{b2eqp2}) implies that
\beqna{b2diff}
 B_1^2-B_0^2 &=& \int_{0}^{x_1}\dadb{B^2}{x} dx   \\
 &=& \int_{0}^{x_1}  4\frac{n}{G} \dadb{P^2}{x} dx= \int_{0}^{x_1} 4\frac{n}{G} \dadb{P^2}{G}\dadb{G}{x} dx  . \nonumber \eeqna
Using the fact $dP^2/dG = -2 G$, and that the constant entropy distribution has 
\[\dadb{G}{x} = \frac{1}{n} \dadb{(nT)}{x},\]
we arrive at (since $n_1=0$)
\beqn{diffB2}
B_1^2-B_0^2  =-8 \int_{0}^{x_1} \dadb{(nT)}{x} dx = 8n_0 T_0.
\eeqn
Thus the difference between the boundary magnetic field and the central magnetic field strength depends only on the central density and temperature, but is independent of $\lambda$, the width of transition, or the orientation of the fields.  This is equivalent to saying that the magnetic pressure and the thermal pressure balance.  

\subsection{Equal $\lambda$: no bulk motion}
If  $\lambda_+=\lambda_-=\lambda$, we can have a solution with no total momentum, $\PP_{tot} = \PP_+-\PP_- =0$, and equations (\ref{eeplus}) and (\ref{eeminus}) become degenerate. Thus 
\beqn{curlp}
 \epsilon \curl \PP=  \epsilon^{-1} \lambda\frac{n}{G} \PP  -\frac{ \BB}{2} , 
\eeqn
where $\PP=\PP_+=-\PP_- $. In 1-D, this reduces to 
\beqn{curlpy}
\epsilon\dadb{P_z}{x} = -\epsilon^{-1}\lambda  \frac{n}{G} P_y + \frac{1}{2}B_y,
\eeqn
\beqn{curlpz}
\epsilon\dadb{P_y}{x} = \epsilon^{-1}\lambda  \frac{n}{G} P_z - \frac{1}{2}B_z.
\eeqn
The current equation (\ref{eeAmpere}) becomes 
\beqn{curlBy}
\epsilon \dadb{B_y}{x} =2\frac{n}{G} P_z,
\eeqn
\beqn{curlBz}
\epsilon \dadb{B_z}{x} =-2\frac{n}{G} P_y.
\eeqn
We note that the 1-D constraint enforces $P_x=B_x=0$.  

In Figure \ref{fig2}, we show two examples of one dimensional solutions with equal $\lambda=1$. Both solutions have $T_0=1,\ n_0=1$, and $P_y(x_0) = 1$. In the first plot (a), the central magnetic field vanishes at the origin. There is a finite region over which the plasma density is supported. The magnetic field changes in both direction and magnitude over a width of $\Delta x = 9.77$. As discussed above, the magnitude of $B$ at the edge is set by the central density and temperature. In this case, $B^2(x_1) = 8.0$, and the field orientation twists through an angle of $\Delta\theta = 131.5^{\circ}$.  In (b), the central field is taken as $B_y(x_0) = 1$. Nowhere does the magnetic field vanish in this case. The transition width is $\Delta x = 8.68$, the edge magnetic field strength is $B^2(x_1) = 9.0$,
and the angle change is  $\Delta\theta = 109.3^{\circ}$. Note also that the direction of $J_z$ is reversed. 
 \begin{figure}[t]
   \centering
   \includegraphics[width=.45 \textwidth]{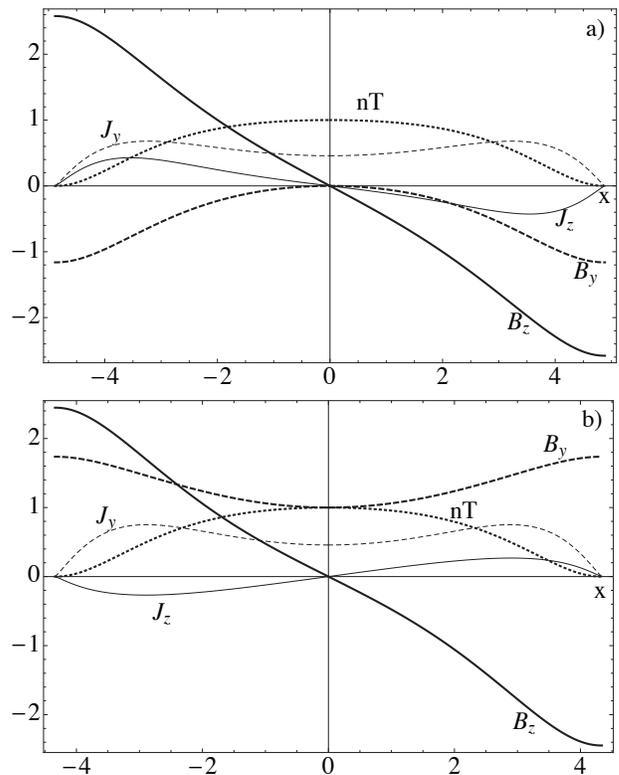} 
   \caption{Two examples of pair-plasma equilibrium state in a 1-D configuration $\lambda_+=\lambda_-=1$. There are components of $\BB$ and $\JJ$ in both directions perpendicular to $\hat{x}$, with a finite transition width dictated by the equilibrium equations and the adiabatic condition. In (a), the magnetic field vanishes at the origin, while it retains a finite value in (b).   }
   \label{fig2}
\end{figure}

\subsubsection{Sheet-Pinch}
With a sheet-pinch type equilibrium in mind, take $\curl\BB=\alpha \BB$, with $\alpha$ a scalar function of $x$. Then inserting  $\PP_+=-\PP_-=\PP$ into equation (\ref{eeAmpere}) shows that
\[ \PP = \epsilon \frac{\alpha G}{2 n} \BB. \]
For this case, $B^2$ and $P^2$ are both constant, which means that $n/G$ and $\alpha$ are both constants as well. The 1-D solution is the sheet pinch $\BB = B_0( \sin(\alpha x)\hat{y}+ \cos(\alpha x)\hat{z} )$.  However, in order to maintain $\PP$ and $\BB$ parallel, we require that
\[ 2 \frac{n}{G}\left(\epsilon \frac{\alpha G}{2 n}\right)^2 +\frac{1}{2} - \epsilon^{-1}\lambda\frac{n}{G}\left(\epsilon \frac{\alpha G}{2 n}\right) =0\]
\[\epsilon^2 \alpha^2 \frac{G}{ n}  -\lambda \alpha  +1=0.\]
Thus the temperature and density determine the scale of the field.
If $\epsilon$ is small (skin depth $<<$ system size), we can take $\alpha= 1/\lambda$, none of the relativistic effects matter, and the $\lambda$ scale is set by the system boundary conditions.  The momentum is small compared with the magnetic field strength in this limit. 

However, if skin depth effects are important, 
\[ \alpha =  \frac{\lambda}{2 \epsilon^2}\frac{n}{G} \left(1 \pm \sqrt{1-4 \epsilon^2 \frac{G}{\lambda^2 n}}\right). \]
Real solutions imply that $ \lambda > 2 \epsilon \sqrt{G/n}$. The fact that one solution diverges as $\epsilon\to 0$ shows the singular perturbation nature of the underlying equations.

\section{Discussion}
We have shown that the evolution equations for charged, relativistic, homentropic fluids can be expressed in a vortex form, using a minimal coupling method having the same form as the  canonical momentum of single particle motion, but with the inclusion of thermal effects. This places the magnetic field and fluid vorticity on the same footing. Following this formalism, we found that the effective Hamiltonian constructed from this canonical momentum leads to a well-posed variational principle, which allowed us to find relaxed equilibrium states of the relativistic equations of motion. These equilibria can exhibit structures on multiple spatial scales, and could be very useful in the modeling of hot astrophysical flows.  We have made no claims as to the stability of the states--indeed we expect the configurations containing counter-streaming species to be unstable in the kinetic regime. However, such states could be useful as starting points for studies of relativistic magnetic reconnection. 

The class of one dimensional solutions has a wide variety of possible configurations depending on the symmetries between the fluids. The multitude of these states must be narrowed down based on the  boundary conditions particular to a given application. One possible application of this type of configuration is the modeling of the `striped wind'  that results in Pulsar Wind Nebulae (PWNe) with unaligned rotating dipoles \cite{1990ApJ...349..538C,1994ApJ...431..397M, 2008AIPC..983..200A}. In this configuration, an expanding entropy wave consists of low entropy regions of  constant magnetic field separated by high entropy, neutral pair-plasma current sheets. A typical PWNe of this type can have $\gamma G\sim 100,$ making a relativistic treatment necessary. It is expected that the dissipation within the magnetic field transition region happens on a much shorter timescale than the expansion of the wave, enabling the application of equilibrium theory. The reversed field configuration described in Section \ref{reversal} provides a detailed and physics-rich description of this transition region, evaluated in the frame moving with the entropy wave.  Although we have presented here solutions with constant entropy inside the hot, dense transition region, this assumption can easily be relaxed in the 1-D case. Outside of the transition regions, the constant field portion of the wave can be described by the case presented in subsection \ref{constant}, and the entropy can be much lower than in the transition region. The magnetic pressure in the constant field regions balance the thermal pressure within the current sheets, as in previous studies. We will further examine the application of the formalism presented in this paper to these configurations in future work. 

\appendix
\section{Conserved Quantities}
Below, we show under what conditions the variational constraints used in the minimization principle are conserved. We use the general conservation principle that if a 4-vector $V^{\nu}$ is divergence free ($\del_{\nu}V^{\nu}=0$), then the spatial integral of the time component is conserved, provided surface terms vanish. The total energy is automatically an ideally conserved quantity because the closure $J^{\nu}=\sum_s q_s \Gamma_s^{\nu} $ implies that $\del_{\mu}T_{tot}^{\mu\nu}=0$. 

To define the total helicity of each species, we take the dual of $M$, $\mathcal{M}^{\mu\nu} = 1/2 \epsilon^{\alpha\beta\mu\nu}M_{\alpha\beta}$, and construct the helicity 4-current
\[ K^{\mu}= \mathcal{M}^{\mu\nu} \Pi_{\nu}. \] 
The divergence of this 4-current is 
\beqna{divK}
\del_{\mu}K^{\mu}&=& \mathcal{M}^{\mu\nu} \del_{\mu}\Pi_{\nu} = \frac{1}{2} \mathcal{M}^{\mu\nu}M_{\mu\nu} \nonumber \\
&=& -2 \bm{\Omega}\cdot (\EE + c/q \QQ),
\eeqna 
where we have used the fact that the divergence is a scalar to evaluate in the lab frame, and we use the same definitions as in Section \ref{sec2}.  From (\ref{evolution}), we have the lab-frame value:
\[ (\EE + c/q \QQ) = -
 \uu \times \OO - \frac{T}{q \gamma}\grad \sigma,\] 
 which allows us to write 
 \beqna{divK2}
\del_{\mu}K^{\mu}&=& 2 \frac{T}{q \gamma} (\bm{\Omega}\cdot \grad) \sigma.
\eeqna 
 Thus we see that helicity will be conserved if the entropy is constant on lines of total vorticity, which is true for isentropic plasmas in the case that $\uu\|\bm{\Omega}$. In the limit of small mass, this implies that entropy is a magnetic flux function, a standard result for MHD.  In addition, we note that if there exists an equation of state such that the thermodynamic term can be expressed as a full gradient, 
 \beqn{eos} (T/\gamma)\grad\sigma = \grad \zeta,\eeqn
  then due to the fact that $\diver\bm{\Omega}=0$, the 4-divergence of the helicity can still be written as a total divergence
  \beqna{divK3}
\del_{\mu}K^{\mu}&=& \frac{2}{q} \bm{\Omega}\cdot \grad \zeta = \diver\left( \frac{2}{q} \bm{\Omega}\zeta \right).
\eeqna 
 Then the species helicity 
\beqn{heldef}
h =\int K^0\ d^3 x = \int [\BB+(c/q)\bm{R}]\cdot[\bm{A}+(c/q)\PP]\ d^3 x, \eeqn
is a conserved quantity. Note that the equation of state condition (\ref{eos}) is identical to the conservation of vorticity condition in equation (\ref{vorticity}); namely that the baroclinic term vanishes. Thus we conclude that whenever the flow preserves total species vorticity $\Omega$, the helicity $h$ is conserved. Note that in the 1-D examples of section \ref{examples}, the baroclinic term always vanishes regardless of the form of entropy. 

This helicity contains a term involving the momentum times the curl of the momentum ($\sim v^2 k$). Thus it has a higher order of spatial derivative than the total energy, which contains only the momentum squared ($\sim v^2$). It is for this reason that the helicity is more fragile than energy to dissipation, and we must use the variational principle (\ref{fullminimize}).

%


\end{document}